\documentclass[conference]{IEEEtran}

\IEEEoverridecommandlockouts
\usepackage{cite}
\usepackage{amsmath,amssymb,amsfonts}
\usepackage{algorithmic}
\usepackage{graphicx}
\usepackage{textcomp}
\usepackage{xcolor}
\usepackage{tablefootnote}
\usepackage{siunitx}
\usepackage{makecell}
\usepackage{booktabs}
\usepackage{todonotes}
\usepackage{algorithm}
\usepackage{url}
\usepackage[acronym]{glossaries}

\makeglossaries

\def\BibTeX{{\rm B\kern-.05em{\sc i\kern-.025em b}\kern-.08em
    T\kern-.1667em\lower.7ex\hbox{E}\kern-.125emX}}
\begin{document}

\newacronym{per}{PER}{Packet Error Rate}
\newacronym{pdr}{PDR}{Packet Delivery Ratio}
\newacronym{rtt}{RTT}{Round Trip Time}
\newacronym{hvac}{HVAC}{Heating, Ventilation, and Air Conditioning}
\newacronym{adas}{ADAS}{Advanced Driver Asistance Systems}
\newacronym{ram}{RAM}{Random Access Memory}
\newacronym{ai}{AI}{Artificial Intelligence}
\newacronym{3gpp}{3GPP}{Third Generation Partnership Projects}
\newacronym{cpu}{CPU}{Cental Processing Unit}
\newacronym{vcpu}{vCPU}{virtual Cental Processing Unit}
\newacronym{gpu}{GPU}{Graphics Processing Unit}
\newacronym{nsa}{NSA}{Non-Standalone}
\newacronym{asil}{ASIL}{Automotive Safety Integrity Level}
\newacronym{iaas}{IaaS}{Infrastruture as a Service}
\newacronym{faas}{FaaS}{Function as a service}
\newacronym{paas}{PaaS}{Platform as a service}
\newacronym{tcp}{TCP}{Transmission Control Protocol}
\newacronym{ecu}{ECU}{Electronic Control Unit}
\newacronym{lte}{LTE}{Long Term Evolution}
\newacronym{rssi}{RSSI}{Received Signal Strength Indicator}
\newacronym{5G}{5G}{Fifth Generation}
\newacronym{embb}{eMBB}{Extended Mobile Broad Band}
\newacronym{mmtc}{mMTC}{massive machine-type communications}
\newacronym{urllc}{uRLLC}{Ultra Reliable and Low Latency Communication}

\newacronym{de-cix}{DE-CIX}{Deutsche Commercial Internet Exchange}

\title{Directives for Function Offloading in 5G Networks Based on a Performance Characteristics Analysis}

\makeatletter
\newcommand{\linebreakand}{%
  \end{@IEEEauthorhalign}
  \hfill\mbox{}\par
  \mbox{}\hfill\begin{@IEEEauthorhalign}
}
\makeatother

\author{
\IEEEauthorblockN{Falk Dettinger, Matthias Weiß and Michael Weyrich}
\IEEEauthorblockA{
\textit{Institute of Industrial Automation and Software (IAS)} \\
\textit{University of Stuttgart} \\
Pfaffenwaldring 47, 70550 Stuttgart, Germany \\
E-Mail: \{dettinger, weiss, weyrich\}@ias.uni-stuttgart.de}
\and
\IEEEauthorblockN{Daniel Baumann and Martin Sommer}
\IEEEauthorblockA{
\textit{Institut fuer Technik der Informationsverarbeitung (ITIV)} \\ 
\textit{Karlsruhe Institute of Technology (KIT)} \\
Engesserstr. 5, 76131 Karlsruhe, Germany \\
E-Mail: \{daniel.baumann, ma.sommer\}@kit.edu}
}


\maketitle

\begin{abstract}
Cloud-based offloading helps address energy consumption and performance challenges in executing resource-intensive vehicle algorithms. Utilizing 5G, with its low latency and high bandwidth, enables seamless vehicle-to-cloud integration. Currently, only non-standalone 5G is publicly available, and real-world applications remain underexplored compared to theoretical studies.
This paper evaluates 5G non-standalone networks for cloud execution of vehicle functions, focusing on latency, \glsentrylong{rtt}, and packet delivery. Tests used two AI-based algorithms—emotion recognition and object recognition—along an 8.8 km route in Baden-Württemberg, Germany, encompassing urban, rural, and forested areas. Two platforms were analyzed: a cloudlet in Frankfurt and a cloud in Mannheim, employing various deployment strategies like conventional applications and containerized  and container-orchestrated setups.
Key findings highlight an average signal quality of 84\,\%, with no connectivity interruptions despite minor drops in built-up areas. Packet analysis revealed a \glsentrylong{per} below 0.1\,\% for both algorithms. Transfer times varied significantly depending on the geographical location and the backend servers' network connections, while processing times were mainly influenced by the computation hardware in use. Additionally, cloud offloading seems only be a suitable option, when a round trip time of more than 150\,ms is possible.
\end{abstract}


\section{Introduction}

The rapid development of connected vehicles has brought about a wave of technological innovation, including the need for advanced data processing and real-time decision-making. Modern vehicles generate vast amounts of data from a wide range of sensors and onboard systems, creating a demand for powerful computing resources that require large onboard processing units. Furthermore, because modern vehicles have a continuously growing number of functions  \cite{Antinyan.2020}, the need for even greater computational power and resource management has become increasingly critical. The resources available within a vehicle are inherently limited because they are fixed and determined during the vehicle design and manufacturing process. These constraints further emphasize the trade-off between computational power and energy efficiency, highlighting the need for efficient resource management in the integration of AI systems \cite{Ammal.2021}. However, AI models often require significant amounts of energy depending on the specific model and its complexity. Thus, this may challenge the performance versus energy consumption in the vehicle, particularly for large computationally intensive models.

Offloading certain algorithms to the cloud to mitigate the drawbacks of local execution such as energy consumption and hardware performance limitations is possible and advantageous. Consequently, outsourcing complex tasks such as \gls{hvac} functions with machine learning or cooperative perception of the cloud has become an increasingly viable solution \cite{BaumannSommerDettinger2024_1000171757}. In addition, this approach can reduce costs by minimizing the need for expensive on-board hardware and taking advantage of the scalability offered by cloud resources \cite{Baumann.2024}. However, the adoption of cloud computing for vehicles introduces challenges, particularly concerning the network infrastructure \cite{falchetti2015vehicular}. Traditional cellular technologies struggle to meet the stringent requirements of real-time vehicle communication, particularly in terms of latency, bandwidth, and reliability. This is when 5G technology plays a transformative role. With the promise of ultra-low latency and high bandwidth, 5G networks are uniquely positioned to support the seamless integration of vehicles with cloud services \cite{arthurs2021taxonomy}. 

This paper addresses the need for mobile connectivity for cloud-offloading applications in the automotive domain. The paper aims to provide insights into the capabilities of 5G networks in overcoming the barriers to cloud offloading and how this technology can drive the next wave of innovation in automotive systems. In addition, it includes an analysis of the 5G network to estimate the achievable communication rates and examines the implications of these rates for the proposed functions that can be enabled through Cloud Computing. Finally, directives are outlined to indicate the conditions under which cloud offloading is suitable.

 


           


\section{Background} 
\subsection{Mobile Communication Networks}

\gls{5G} mobile technology offers speeds of up to 10 Gbps and ultra-low latency of 1\,ms at its minimum \cite{hakak2023autonomous}. As such, it is significantly more powerful than \gls{lte}, particularly since it supports \gls{embb}, \gls{mmtc}) and \gls{urllc}, which are essential for enabling a wide range of use cases in vehicular environments \cite{hakak2023autonomous}.

Although \gls{lte} networks are already highly advanced, transitioning from \gls{lte} to \gls{5G} necessitates different hardware. This makes rapid upgrade inefficient due to a lack of immediate use cases and high costs \cite{moreno20245g}. To address this, an intermediate approach known as \gls{5G} \gls{nsa} was introduced. \gls{5G} \gls{nsa} uses the \gls{5G} radio data plane while continuing to rely on the \gls{lte} control plane and \gls{lte} core \cite{alnaas11comparison}.


\subsection{Evaluation Metrics}\label{sec:metrics}
In the literature, several metrics are used to classify the performance of communication networks. In the following, the most common are summarized according to definitions provided in the literature.
\begin{itemize}
    \item \textbf{Signal Quality:} The signal quality describes the connection quality of the 5G module or the signal strength received in percent. There are multiple metrics by which this percentage can derived, the most common being the \gls{rssi}, which measures the power of the signal at the receiver relative to the maximum receivable power \cite{witrisal20212signalstrength}.
    \item \textbf{Latency:} Latency describes the transmission delay \cite{mannoni2019latency}. In this specific case, the latency encompasses both upload and download latencies.
    \item \textbf{Processing Time:} The processing time indicates the time required to process the input data and provide the results as variable. It is directly affected by hardware performance. \cite{harris2021understanding}
    \item \textbf{\gls{rtt}:} The \glsentrylong{rtt} is defined as the time it takes to send a request plus the time to receive back an acknowledgment \cite{afzal2023rtt}. For this paper, this definition is expanded to also include the processing time of the request on the server and the time it takes until the response has reached the client. Thus, it can be described as the total time it takes until a client request is completed.
    \item \textbf{\gls{per}:} The \glsentrylong{per} indicates the number of incorrectly received or not received packets \cite{pundir2021per}. Depending on the transmission protocol, a high PER prolongs the transmission duration or reduces the quality of the result because packets need to be retransmitted or unavailable due to packet loss.
    \item \textbf{\gls{pdr}:} The \glsentrylong{pdr} compares the number of correctly received packets against the total number \cite{dettinger2024survey}.
\end{itemize}

\section{Related Work}
It can be observed that recent literature increasingly focuses on the offloading of connected vehicle functions, where both the implementation of the offloading processes itself and analyzing the requirements to perform function offloading are focused. For example, the authors in \cite{ashok2018vehicular} have shown that non-time-critical vehicle functions can be outsourced to the cloud. With regard to time-critical functions, the authors observe that high processing times and \gls{rtt}s overall have a significant impact on the feasibility, concluding that partial function offloading or improved network performance are required. 

As similar observations become more common, publications try to design generalizable frameworks by which the feasibility of function offloading for certain functions can be determined. Following this trend, previous work by the authors also focused on identifying vehicle functions for offloading \cite{10.1007/978-3-031-62277-9_38                }. Here, the authors proposed a two-step process to determine both feasibility and suitability. Feasibility is based on the safety criticality of the function. Therefore, the authors proposed that functions with an assignment of an \gls{asil} are not feasible for cloud offloading. The second step of the process assigns a suitability score to each function by considering factors like power consumption, hardware cost and potential benefits by cloud execution.

Yet, even if a function is judged to be suitable for cloud offloading by applying such a process, being able to perform the offloading during operation is heavily influenced by the available infrastructure and current network conditions. For example, Boutin et al. \cite{boutin2021edge} performed function offloading for image processing inside an indoor 5G network. While showing the effectiveness of the approach under ideal network conditions, results also indicate a quick deterioration when the process is disturbed by the environment. A similar observation was made by Xiong et al. \cite{xiong2021v2x}, who concluded that offloading reliability is highly dependent on network quality. In their approach, they derive an upper bound for network delay using stochastic methods and, based on this, calculate failure probabilities for different offloading techniques, which can then be considered in the optimization process. Their simulation setup  highlights clearly that a) the probability of failing the offloading increases under difficult network conditions and b) by considering network parameters in the offloading process, more failures can be avoided.

To summarize: While many publications prove the feasibility of their approaches to offload certain functions, applying offloading to a real-world scenario often leads to limited results or requires advanced mitigation techniques that consider the network's performance characteristics. Despite these indications, the authors of this work are not aware about any popular set of rules to guide the process, a gap which this paper aims to close.

%
%

%

\begin{figure}[t]
    \centering
    \includegraphics[width=1.0\linewidth]{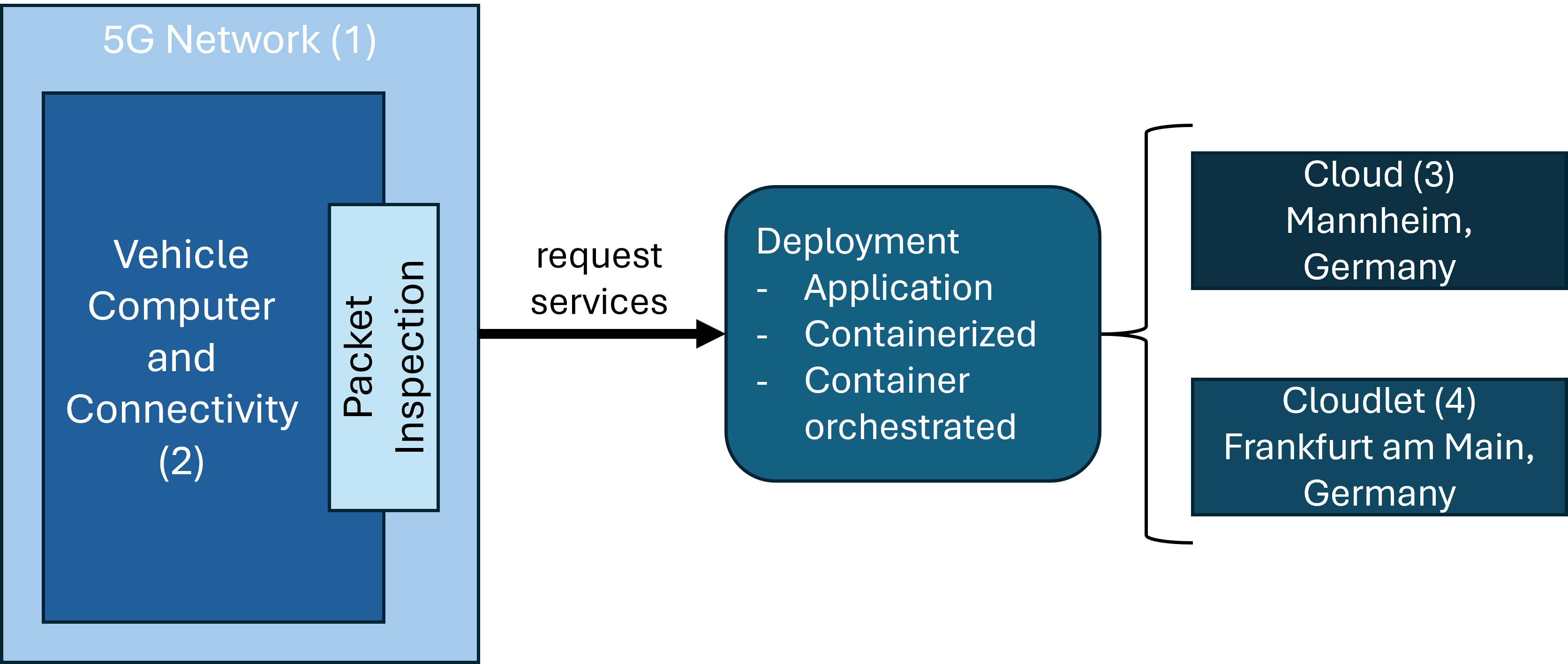}
    \caption{Overview of the experimental setup. The vehicle is connected to the 5G network and executes functions that are deployed using different deployment strategies on the cloud or the cloudlet. The network traffic is monitored using Wireshark.}
    \label{fig:Evaluation-setup}
\end{figure}

\section{Evaluation}
In this chapter, it is investigated whether two sample functions can successfully be offloaded to the backend under common network conditions in Germany. Performance characteristics of the network are measured during the experiments. Based on these results, directives for function offloading are provided in Section \ref{sec:discussion}.

\subsection{Evaluation Setup}
Fig. \ref{fig:Evaluation-setup} summarizes the component overview and the structure of the evaluation setup. The enumeration in the figure is also used in the following paragraph.

\begin{enumerate}
    \item \textbf{Vehicle Computer and Connectivity}: The vehicle computer is an HP Victus16-r1174ng laptop with an external 5G module 5G M2 EVB-Kit \cite{Quectel5Gboard} manufactured by Quectel, equipped with a 5G M.2 module of the type RM520N-GL \cite{Quectel5Gmodule}. This supports not only 5G but also 4G and 3G in multi-mode as well as 5G NR according to \gls{3gpp} Release 16 \cite{dettinger2024survey} in 5G non-standalone and 5G standalone. The TCP protocol is used for communication between the vehicle and the cloud, which ensures sufficient packet capture on a system node. The communication with the cloud, i.e., all incoming and outgoing packets, is captured using Wireshark and subsequently analyzed.
    \item \textbf{5G Network:} The 5G \gls{nsa} network of Deutsche Telekom is used, with a dedicated network slice providing a guaranteed bandwidth of at least 100 Mbps.
    \item \textbf{BWCloud:} This cloud is an \gls{iaas} platform hosted by the state of Baden-Württemberg, located in Mannheim, Germany. Functions are executed on the IaaS platform as conventional applications, containerized in a Docker container, as well as container-orchestrated in a Kubernetes cluster. The cloud instances used have two \gls{vcpu}, four GB of \gls{ram}, and 60 GB of disk capacity.
    \item \textbf{Telekom Cloudlet:} The \gls{faas} cloudlet platform is provided by Deutsche Telekom. The location is in Frankfurt am Main, Germany. Due to its execution as a \gls{paas} platform, only functions that are executed within a container and container-orchestrated can be run in the cloudlet. The instances used have two \gls{vcpu}, four GB of \gls{ram}, and 40 GB of disk capacity.
\end{enumerate}

The metrics introduced in Section \ref{sec:metrics} are used to evaluate the network performance. For evaluation, two example algorithms are employed. Test algorithm 1 utilizes an emotion recognition algorithm that cyclically detects the driver's emotions at a frequency of 2 Hz. It is based on Google Mediapipe with pre-trained weights. Whereas, test algorithm 2 employs the object recognition algorithm YOLOv8 to process fundamental environmental information, such as vehicles, traffic signs, and pedestrians. Safety-critical applications require a frequency of 50 Hz (20 ms) \cite{protzmann2019large}. In our case it will be processed with 4 Hz up to 6 Hz. The algorithm is based on pre-trained YOLOv8n weights.

The functions for evaluating emotions and object recognition follow the same process flow. The vehicle establishes a \gls{tcp} connection to the service and periodically sends input data once the connection is established. The service processes the input data, measures the execution time for emotion or object recognition, and returns both the result and execution time to the vehicle. Inside the vehicle, details about the input and output data of the service, along with \gls{rtt}, processing time, and 5G network signal strength, are stored in a log file. The vehicle initially connects to the backend server service and performs the functions as long as necessary. 
In parallel to execution, communication between the vehicle and the cloud is captured using Wireshark.

\begin{figure}[b]
    \centering
    \includegraphics[width=\linewidth]{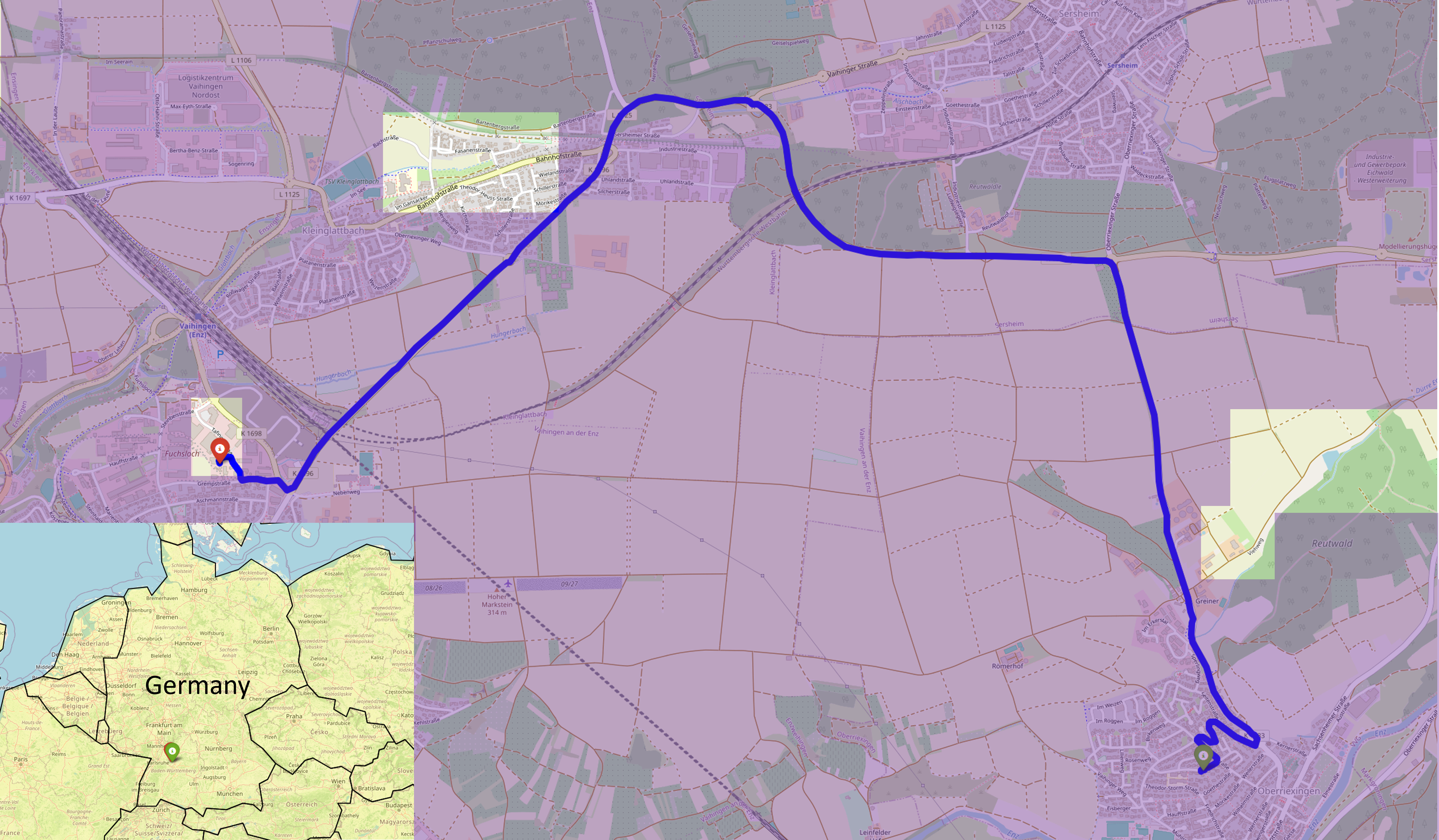}
    \caption{Map outlining the route in southern Germany including the 5G \glsentrylong{nsa} coverage in purple.}
    \label{fig:map}
\end{figure}

All measurements were carried out along an 8.8\,km route over a state road in Baden-Württemberg, Germany. Depending on the traffic density, the driving duration was 10 to 12\,minutes. The region is comprehensively covered by 5G \gls{nsa} and connects both rural and urban areas. Furthermore, the route passes through open spaces as well as forested and developed areas (see Fig. \ref{fig:map}). The purple area outlines the 5G coverage provided by the telecommunication provider. During the drive, the object recognition algorithm was executed 1,770 times, while the emotion recognition algorithm was executed 1,010 times. For the emotion recognition algorithm, a set of ten AI-generated input images with an average size of 147.31\,kBs. With an average execution time of 350\,ms and an average data size of 420.9\,kB are transferred per second. For the object recognition algorithm, ten input images with an average size of 222.8\,kBs were used. With an average processing time of 240\,ms, an average data size of 833\,kB/s is transferred.
Throughout the driving duration, network traffic on the vehicle side was captured using Wireshark, and packet flow was analyzed using packet analysis regarding \gls{per} and \gls{pdr}. Fig. \ref{fig:Evaluation-setup} illustrates the evaluation setup.
Both functions are executed on the \gls{iaas} platform as conventional applications, containerized in a Docker container, and container-orchestrated in a Kubernetes cluster. On the \gls{faas} platform the functions can only be executed in a containerized and container-orchestrated manner.

\begin{figure*}[t]
    \centering
    \includegraphics[width=\linewidth]{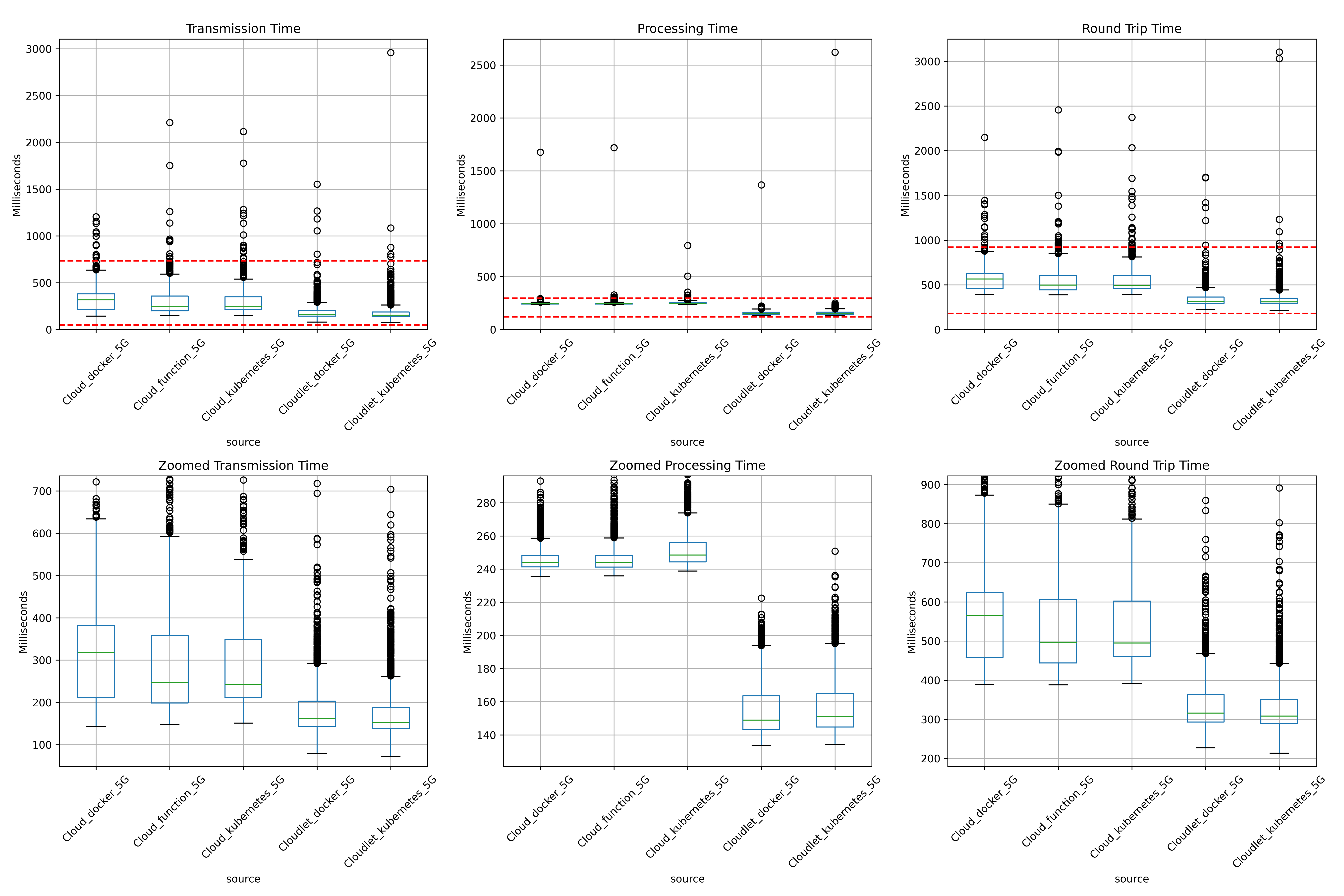}
    \caption{Box plots for executing the object recognition algorithm in the BWCloud in Mannheim and the Telekom Cloudlet in Frankfurt. The functions are executed as conventional functions, containerized using Docker and container orchestrated using Kubernetes.}
    \label{fig:boxplot_object}
\end{figure*}

\begin{figure*}[t]
    \centering
    \includegraphics[width=\linewidth]{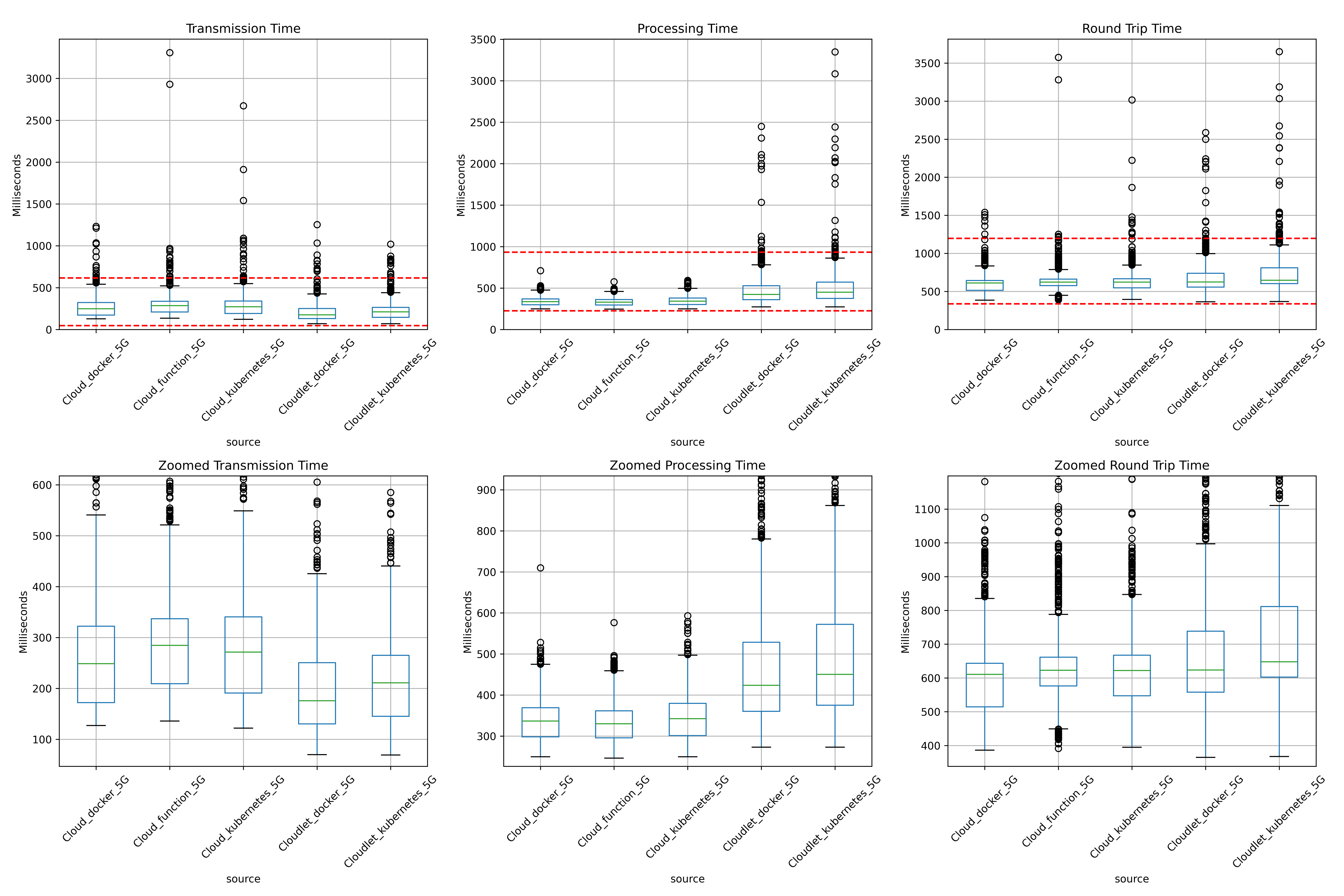}
    \caption{Box plots for executing the emotion recognition algorithm in the BWCloud in Mannheim and the Telekom Cloudlet in Frankfurt. The functions are executed as conventional functions, containerized using Docker and container orchestrated using Kubernetes. The zoomed out are is limited by the red dash-dotted line.}
    \label{fig:boxplot_emotion}
\end{figure*}

\subsection{Results}\label{sec:result}
The evaluation of \textit{signal quality} shows that the average mean is around 84,54\,\% during all 8,850 function calls of the object recognition algorithm, while the minimum signal quality measured was 50\,\%. For the 5,050 calls of the emotion recognition algorithm, the signal qualities average mean is 84.58\,\% with a minimum signal quality of 52\,\%. Especially in villages and cities where buildings interfere with the radio connection, the signal quality drops slightly. No complete connection interruption occurred, as there are no tunnels on the chosen route that block radio connections.

For the evaluation of the \gls{per} and the \gls{pdr}, packets transmitted by the vehicle computer are considered. This includes the transmitted payload data, and also acknowledgment packets, retransmitted packets, duplicates, and fast retransmissions.

The analysis of the data packets captured using Wireshark show that an average of 395,271 data packets are exchanged between the vehicle computer and the cloud per test drive for the object recognition algorithm. Of the total exchanged packets, an average of 283,151 is sent from the vehicle computer to the cloud, while an average of 88.8 duplicate packets are captured. Thus, the average \gls{per} can be specified as 0.031\,\% and the average \gls{pdr} as 99.969\,\%. 

In contrast, an average of 155,501 data packets are transferred between the vehicle and the network server for the emotion recognition algorithm. On average, 108,710 packets are sent from the vehicle computer to the cloud, while an average of 72.2 duplicate packets are captured. Thus, the resulting \gls{per} is 0.066\,\% and the \gls{pdr} 99.94\,\%.

From the measurement data of the signal quality, it is apparent that it typically lies in an average range of over 80\,\%, and a quality below 50\,\% is rare in well-developed regions. Furthermore, the packet traces indicate that the connection quality in the measured areas has no impact on the \gls{per} and thus the reliability of the transmission.

Fig. \ref{fig:boxplot_object} and Fig. \ref{fig:boxplot_emotion} sum up the transmission time, the processing time, and their sum given as \gls{rtt} for the deployment of the algorithms grouped into the \gls{iaas} platform, indicated with \textit{Cloud} and the \gls{faas} platform indicated with \textit{Cloudlet}. The extension \textit{\_docker}, \textit{\_function}, and \textit{\_kubernetes} indicate the type of deployment.

Considering the \textit{object recognition-services'} \textit{\gls{rtt}} (see Fig. \ref{fig:boxplot_object}, upper right), it can be seen that the overall duration is located between 200\,ms and around 3 seconds. 
It is evident that the \gls{faas} platform provides results faster than it is possible for the \gls{iaas} platform. This is also illustrated in Fig. \ref{fig:boxplot_object} at the bottom right. In this figure, it can be observed that the total time until the result is available again at the sender is approximately twice as high when using the cloud in Mannheim compared to using the cloudlet in Frankfurt.
Considering the \textit{Processing Time} (see Fig. \ref{fig:boxplot_object} middle), it becomes apparent that the processing times, with approximately 250\,ms for the cloud and 150\,ms for the cloudlet, differ significantly, even though the instances have the same metrics in general. It is also clear that outliers larger than 300\,ms rarely occur, but when they do, they are significantly larger than the processing time itself.
The median of the \textit{Transmission Time} in the conducted tests ranges from approximately 150\,ms to 170\,ms when using the cloudlet, and from 240\,ms to 310\,ms when using the cloud. In general, the transmission time to the cloud is higher than to the cloudlet, even though the distances are of the same order of magnitude. In general, the measured outliers are in a similar range. In advance, it can be seen that the spread of measured values on each device is independent of the deployment strategy \textit{conventional function}, \textit{containerized}, or \textit{container orchestrated}.

In contrast to the measurement of the \textit{object recognition service} in \ref{fig:boxplot_object}, the results of the \textit{emotion recognition service} are summarized in Fig. \ref{fig:boxplot_emotion}.

For the \textit{\gls{rtt}} (Fig. \ref{fig:boxplot_emotion}, right), there is a trend of higher overall time on the cloudlet, with the median across both devices, for all deployment strategies, being in the range of about 610\,ms to 650\,ms. A major portion of the outliers appears for the \gls{iaas} platform in the time range up to 1 second, while the outliers for the \gls{faas} platform appear in the range of 1.2 seconds to 1.5 seconds. Comparing the \textit{Processing Time}, it is noticeable that the algorithm's processing duration on the \gls{faas} cloudlet is higher than on the \gls{iaas} cloud. While the processing time on the cloud is limited to a range of about 250\,ms to 700\,ms, the processing time on the cloudlet is roughly 270\,ms to 3.5 seconds. The \textit{Transmission Time} to the cloud has a median between approximately 250\,ms and 290\,ms, which is higher than that of the cloudlet. There, the median transmission time is in the range of about 180\,ms to 210\,ms.

For completeness, the object recognition service, deployed as a container on the \gls{iaas} platform and the \gls{faas} platform (results in (brackets)), is also examined along the route using \gls{lte}. It becomes apparent that the signal values read by the communication module average at 62.0\,\% (59.0\,\%), while the \gls{rtt} drops to a median of 480\,ms (330\,ms). The processing time remains at approximately 250\,ms (150\,ms), whereas the transmission time  decreases to 240\,ms (170\,ms). The same behavior can also be seen when considering the emotion recognition service, deployed as a container on both the platforms. Additionally, the transmission time is highly dependent from the data size to be transferred between client and backend server.

\section{Discussion}\label{sec:discussion}
In the following, the results measured in Section \ref{sec:result} are discussed, and the main findings are summarized. Furthermore, based on these results, directives for enabling cloud offloading are elaborated.
In summary, the measurements indicate that the average signal quality exceeds 80\,\%, demonstrating a robust communication connection along the route. While a few measurements show signal quality below 50\,\%, this suggests that a stable radio connection generally prevails in the measured areas. However, shielding caused by nearby buildings does lead to a decrease in signal quality, although it does not affect the \gls{per}. Additionally, the analysis of the \gls{per} reveals an exceptionally low error rate ($<$ 0.07\,\%), resulting in a \glsentrylong{pdr} of nearly 100\,\%. The absence of tunnels and other obstacles, such as tall buildings, along the chosen route may have contributed to these stable connection results.

The analysis of the two algorithms shows that the data transmission time to the cloudlet in Frankfurt am Main is significantly lower than to the cloud in Mannheim (between 150\,ms up to 210\,ms compared to 240\,ms up to 310\,ms). Since the route is the same in all cases and Mannheim is only about 75\,km away (air-line distance), while Frankfurt is \mbox{137\,km} away (air-line distance), both the transmission distance and the geographical distance can be excluded as direct factors. Based on the time values it can be assumed that the cloudlet's connection is significantly more powerful than the cloud's connection in Mannheim. Since the \gls{de-cix} \footnote{\url{https://www.de-cix.net/en/locations/frankfurt}} node is located in Frankfurt, it is likely that the cloudlet connection can be made with a lower number of hops. It can also be concluded that fluctuations in transmission time arise from varying network loads at the time of testing.

On the other hand, the consideration of processing time shows that the object recognition service is executed about 66\,\% faster on the cloudlet, whereas the emotion recognition service is executed about 30\,\% slower on the cloudlet. Despite identical flavors, specific variations in hardware appear to have a significant impact on processing time. Differences in deployment strategies (conventional function, containerized, container orchestrated) do not significantly affect the measured times according to our data. Additionally, our results indicate that the transmission time may increase when a larger size of data is transferred.

Based on the results shown in Section \ref{sec:result} it can be concluded that, based on the used computation hardware, cloud offloading is suitable but only for specific services. For services providing real time functionality, cloud offloading is definitely not suitable, since the transmission time is up to 310\,ms high at around 200\,kB of data. Services providing nearly real time functionality can be offloaded to cloud servers, in the case a small number of packets is transferred between client and network backend and a minimum round trip time of several hundred milliseconds is acceptable.
Additionally, a direct comparison between \gls{lte} and 5G \gls{nsa} clearly shows that 5G does not always provide better performance than \gls{lte}. In this case, the suspected cause is differed routing between \gls{lte} and 5G within the network, leading to varying latencies.

From the measurements, directives for shifting vehicle functions to the network backend can be established: 
\begin{enumerate} 
    \item \textbf{Hardware and Computational Resources:} The hardware platform used has a direct impact on processing time. This must be taken into account, especially for virtual systems where the technical specifications are not always known. For time-critical cases, specialized hardware such as \gls{gpu} should be used for floating-point calculations. 
    \item \textbf{Network Infrastructure and Server Localization:} Backend servers with a good network connection should be selected to minimize communication time. Geographic proximity alone is not decisive. In this context, it must be taken into account that the function may need to be dynamically migrated to other cloud servers for wide availability. 
    \item \textbf{Data Transfer Efficiency:} The time taken for data transfer highly depends on the data size transferred by the service. To enable efficient cloud offloading, the data size should be as small as possible.
    \item \textbf{Application Deployment Optimization:} The type of deployment strategy (conventional functions, containerized, container orchestrated) has no significant impact on the measured results. Therefore, the deployment strategy can be chosen based on other criteria, such as maintainability and scalability. 
    \item \textbf{Network Stability and Data Integrity:} Rely on robust network infrastructures to ensure high reliability of data transmission. Since results may vary in different situations, connection quality and network status should always be monitored. 
\end{enumerate}

\section{Conclusion}
This paper explores the potential of 5G NSA communication in enabling cloud offloading of vehicle functions. To evaluate this, two use cases, object recognition and emotion recognition, are implemented using various deployment strategies on both a cloud and a cloudlet. Along a predefined route, these functions are accessed via 5G NSA and LTE networks. Key performance metrics, \gls{rtt}, processing time, transmission time, signal quality, and packet traces, are recorded and analyzed. This study demonstrates that cloud offloading is a viable option for specific services, depending on both network and processing conditions. The findings highlight several critical aspects and can be summarized as follows:
\begin{itemize}
    \item The type of deployment, whether 'conventional function,' 'containerized,' or 'container-orchestrated, does not significantly influence computation time
    \item Processing time is directly influenced by the underlying hardware, with performance variations arising based on the specific function and hardware utilized. 
    \item Transmission duration depends on both the connection of the backend server and the prevailing network conditions. Servers with robust connections typically ensure shorter transfer durations, though fluctuations occur based on the time of day and week.
    \item The directives established from the measurements provide a structured framework to identify the conditions and strategies under which cloud offloading is an effective and efficient solution for vehicular functionalities.
\end{itemize}

The analysis of public 5G \gls{nsa} reveals that 5G \gls{nsa} is not always faster than \gls{lte}. In addition to the current network load, the backend server's connection and the routing of packets play a crucial role. Based on the measured values, it can be concluded that significantly more efficient and robust transmission networks must be provided to enable the effective migration of vehicle functions to the cloud. Moreover, the servers must be equipped with specialized hardware to meet the short execution times of vehicular functions while accounting for transmission durations. Therefor, our findings suggest to use cloud offloading in vehicular environments, when assuming instantaneous processing, for functions with a \gls{rtt} of at least 200 ms, as the measured transmission time using \gls{lte} is around 150 ms. Additionally, when using 5G \gls{nsa}, the \gls{rtt} should be at least 300 ms.

\section*{Acknowledgment}
The authors would like to thank the German Federal Ministry of Education and Research (BMBF) (under Grant Number: 16MEE0472) and the Chips Joint Undertaking for the financial support under Grant Agreement No: 101139789 (HAL4SDV). The responsibility for the content of this publication lies with the authors.

\bibliographystyle{IEEEtran}
\bibliography{bib}

\end{document}